\documentclass[preprint]{elsarticle}
\usepackage{graphicx}
\usepackage{amssymb}
\journal{Nucl. Instr. and Meth. A}

\begin{document}

\begin{frontmatter}

%% Title, authors and addresses

%% use the tnoteref command within \title for footnotes;
%% use the tnotetext command for theassociated footnote;
%% use the fnref command within \author or \address for footnotes;
%% use the fntext command for theassociated footnote;
%% use the corref command within \author for corresponding author footnotes;
%% use the cortext command for theassociated footnote;
%% use the ead command for the email address,
%% and the form \ead[url] for the home page:
%% \title{Title\tnoteref{label1}}
%% \tnotetext[label1]{}
%% \author{Name\corref{cor1}\fnref{label2}}
%% \ead{email address}
%% \ead[url]{home page}
%% \fntext[label2]{}
%% \cortext[cor1]{}
%% \address{Address\fnref{label3}}
%% \fntext[label3]{}

\title{Performance of Multi-Pixel Photon Counters for the T2K near detectors}

%% use optional labels to link authors explicitly to addresses:
%% \author[label1,label2]{}
%% \address[label1]{}
%% \address[label2]{}

\author[TOKYO]{M.~Yokoyama}
\cortext[cor1]{Corresponding author. Tel:+81-3-5841-1022, Fax:+81-3-5841-1022.}
\ead{masashi@phys.s.u-tokyo.ac.jp}
\author[KYOTO]{A.~Minamino}
%\ead{minamino@scphys.kyoto-u.ac.jp}
\author[KYOTO]{S.~Gomi}
\author[KYOTO]{K.~Ieki}
\author[KYOTO]{N.~Nagai}
\author[KYOTO]{T.~Nakaya}
\author[KYOTO]{K.~Nitta}
\author[KYOTO]{D.~Orme}
\author[KYOTO]{M.~Otani}
\author[KEK]{T.~Murakami}
\author[KEK]{T.~Nakadaira}
\author[KEK]{M.~Tanaka}

\address[TOKYO]{Department of Physics, University of Tokyo, Tokyo, 113-0033 Japan}
\address[KYOTO]{Department of Physics, Kyoto University, Kyoto, 606-8502 Japan}
\address[KEK]{IPNS, High Energy Accelerator Research Organization (KEK), Tsukuba, Ibaraki, 305-0801 Japan}

\begin{abstract}
%% Text of abstract
We have developed a Multi-Pixel Photon Counter (MPPC) for the neutrino detectors of T2K experiment.
About 64,000 MPPCs have been produced and tested in about a year.
In order to characterize a large number of MPPCs, we have developed a system that simultaneously measures 64 MPPCs with various bias voltage and temperature.
The performance of MPPCs are found to satisfy the requirement of T2K experiment.
In this paper, we present the performance of 17,686 MPPCs measured at Kyoto University.
\end{abstract}

\begin{keyword}
%% keywords here, in the form: keyword \sep keyword
Multi-Pixel Photon Counter \sep Geiger-mode APD \sep photodetector \sep characterization \sep quality assurance
%% PACS codes here, in the form: \PACS code \sep code
\PACS 29.40.Wk
%% MSC codes here, in the form: \MSC code \sep code
%% or \MSC[2008] code \sep code (2000 is the default)

\end{keyword}

\end{frontmatter}

\def\degC{\kern-.2em\r{}\kern-.3em C}
\newcommand{\Vbd}{\ensuremath{V_\mathrm{bd}}}
\newcommand{\delV}{\ensuremath{\Delta V}}
%\newcommand{\delV}{\ensuremath{V_\mathrm{over}}}

%\tableofcontents

%\linenumbers

%% main text
\section{Introduction}
T2K (Tokai-to-Kamioka)~\cite{T2K} is a long baseline neutrino oscillation experiment.
In T2K, an intense muon neutrino beam is produced with a proton synchrotron in J-PARC facility and  sent to the massive Super-Kamiokande detector 295~km away.
The main goals of T2K are a sensitive search for the $\nu_e$ appearance from $\nu_\mu$, 
which is related to the yet-unmeasured neutrino mixing angle $\theta_{13}$, and precise measurements of neutrino oscillation parameters $\Delta m^2_{23}$ and $\theta_{23}$.
In order to achieve the aimed precision, good understanding of the neutrino beam properties and neutrino-nucleus interaction are indispensable.
The near detector (ND280) complex is placed at about 280~m from the proton target to provide this information.

The T2K-ND280~\cite{T2K-ND280} consists of several sub-detectors with specific and complimentary functions.
As the basic elements for the particle detection, most of detectors use the plastic scintillator read out via wavelength shifting (WLS) fibers.
This is a widely used technique, especially in recent accelerator neutrino experiments~\cite{sci-wls}.
In those experiments, multi-anode PMTs (MAPMTs) have been used as the photosensor.
For T2K, MAPMT is not suitable because some of the detectors have to operate under a magnetic field of 0.2~T provided by a dipole magnet originally built for UA1 experiment at CERN, and also because the available space is very limited. 

We selected the Multi-Pixel Photon Counter (MPPC\footnote{MPPC is a trademark of Hamamatsu Photonics.}) as the photosensor for ND280  and started the development in cooperation with Hamamatsu Photonics and KEK Detector Technology Project~\cite{MPPC}.
For ND280, 64,000 MPPCs are used in total.
After three years of study, we developed an MPPC that satisfied our requirements and started the mass production in February 2008.
Because it was the first time that the MPPCs are used on such a large scale, it was necessary to develop a method to characterize each device for quality assurance purpose.
In order to characterize a large number of MPPCs, we have developed a system that simultaneously measures 64 MPPCs at Kyoto University.
Using this test system, we have measured 17,686 MPPCs used for two sub-detectors of ND280, called INGRID~\cite{INGRID} and FGD~\cite{FGD}.

In this paper, the design of the test system, testing procedure and summary of measured performances of T2K-MPPC are presented.

\section{Multi-Pixel Photon Counter for T2K Near Detectors}

\subsection{Multi-Pixel Photon Counter (MPPC)}
The Multi-Pixel Photon Counter (MPPC) is a new photodetector manufactured
by Hamamatsu Photonics, Japan~\cite{HPK}.  An MPPC consists of many
(100 to $>$1000) small avalanche photodiodes (APDs), each with an area of either 25$\times$25, 50$\times$50 or 100$\times$100~$\mu$m$^2$, in an area of 1--9~mm$^2$.

Each APD micropixel independently works in limited Geiger mode
with an applied voltage a few volts above the breakdown voltage (\Vbd).
When a photoelectron is produced, it induces a Geiger avalanche.
The avalanche is passively quenched by a resistor untegrated to each pixel.
The output charge $Q$ from a single pixel
 is independent of the number of produced photoelectrons within the pixel, and
can be written as
\begin{equation}
Q = C (V-\Vbd) \equiv C \delV, \label{eq:gain-Vdep}
\end{equation}
where $V$ is the applied voltage and $C$ is the capacitance of the pixel.
The overvoltage, $\delV \equiv V-\Vbd$ is the parameter that controls the performance of MPPC as we will see later.
It is known that $\Vbd$ is dependent on the temperature with a coefficient of $\sim 50$~mV/K~\cite{Otono:2006zz}.
Combining the output from all the pixels, the total charge from an MPPC is
quantized to multiples of $Q$ and proportional to the number of pixels that underwent Geiger discharge (``\textit{fired}'').
The number of fired pixels is proportional to the number of injected photons if the number of photons is small compared to the total number of pixels.
Thus, the MPPC has an excellent photon counting capability as long as the number of photo-electron does not approach the total number of pixels in the device.

For an MPPC, the operating voltage $V$ is a few volts above the breakdown voltage
and well below 100~V.
The pixel capacitance $C$ is on the order of 10--100~fF, giving a gain of 10$^5$--10$^6$.
These features enable us to read out the signal from the MPPC with simple electronics.
In addition, because the thickness of the amplification region is a few $\mu$m, 
an MPPC is insensitive to the magnetic field and has a fast response.

The photon detection efficiency (PDE) of an MPPC is expressed as a product of three effects:
\begin{equation}
\mathrm{PDE} = \varepsilon_\mathrm{geom} \times \mathrm{QE} \times \varepsilon_\mathrm{Geiger}.
\end{equation}
The geometrical efficiency $\varepsilon_\mathrm{geom}$ represents the fraction of active area in
a micropixel.
Based on a measurement~\cite{taguchi-mthesis}, $\varepsilon_\mathrm{geom}$ is about 0.6 for an MPPC with 50$\times$50~$\mu$m pixel size.
The quantum efficiency of the APD, QE, depends on the wavelength of photon and is typically 0.7--0.8 for the wavelength of 400--500~nm, the range of current interest.
The probability of inducing a Geiger discharge when a photoelectron is generated, $\varepsilon_\mathrm{Geiger}$, depends on $\delV$.
The last factor introduces the $\delV$ dependence of PDE.

The dark noise of MPPC at the room temperature is dominated by the Geiger discharge induced by thermally generated electron-hole pairs.
Because it is amplified with the identical process, a dark noise pulse cannot be distinguished from the signal induced by the external photon irradiation, although the dark noise is mainly at the single photoelectron level in absence of the cross-talk effect described below.
The dark noise rate of MPPC is proportional to the area.
For 50~$\mu$m pixel type MPPC, the dark noise rate per active area is about 300--500~kHz/mm$^2$ at 25\degC\ with a gain of 7.5$\times 10^5$.

There are two known processes which give an additional charge to the original signal; the cross-talk between neighboring micropixels and afterpulse.
The origin of the cross-talk is presumed to be optical photons emitted during avalanche~\cite{crosstalk-ref}
which enter neighboring micropixels and trigger another Geiger discharge.
It gives an additional charge output at the same time as the original Geiger discharge.
The afterpulse of MPPC is considered to be due to delayed release of carriers trapped at lattice defects, giving a time-correlated but delayed charge output to the original signal~\cite{Oide}.

In order to evaluate the MPPC performance, it is important to measure key parameters such as $\Vbd$, PDE, dark noise rate, cross-talk and afterpulse probabilities together with their dependence on the temperature and applied voltage.

\subsection{MPPC for the T2K-ND280}
\begin{figure}[tbp]
\begin{center}
\includegraphics[width=0.4\textwidth]{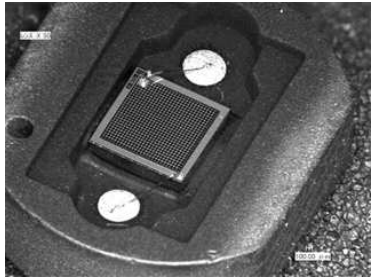}
\caption{Picture of MPPC developed for T2K.}
\label{fig:MPPC}
\end{center}
\end{figure}

A picture of the MPPC for the T2K-ND280 (S10362-13-050C) is shown in Fig.~\ref{fig:MPPC} and
the major specifications is summarized in Table~\ref{tab:spec}.
We use the 1.0~mm diameter Kuraray Y11(200)MS WLS fiber for the ND280 detector.
The sensitive area of the MPPC is enlarged from 1$\times$1~mm$^2$ of those on catalogue (S10362-11-050C) to 1.3$\times$1.3~mm$^2$, so that we can minimize the light loss at the optical contact with a simple coupler.
The size of APD pixel is 50$\times$50~$\mu$m$^2$ and the number of APD pixels~\footnote{In order to make the MPPC fit inside the package, one of the bonding pads needs to be located at the corner of the otherwise sensitive area. This reduces the number of pixels by nine from 26$^2$.} is 667.
The large PDE of MPPC results in a light yield sufficiently large to safely reject the dark noise, which is mainly at a single photoelectron level.
In addition, thanks to the pulsed neutrino beam timing, the random dark noise has only little effect for the neutrino event reconstruction.

\begin{table}[tbp]
\begin{center}
\caption{Specifications of T2K-MPPC(S10362-13-050C).}
\begin{tabular}{ccc} \hline \hline
\multicolumn{2}{c}{Item} & Spec. \\ \hline \hline
\multicolumn{2}{c}{Active area} & 1.3$\times$1.3~mm$^2$ \\ \hline
\multicolumn{2}{c}{Pixel size} & 50$\times$50~$\mu$m$^2$ \\ \hline
\multicolumn{2}{c}{Number of pixels} & 667 \\ \hline
\multicolumn{2}{c}{Operation voltage} & 70~V (typ.) \\ \hline
\multicolumn{2}{c}{PDE @ 550~nm} & $>$15\% \\ \hline
Dark count &($>$0.5~pe) & $<$1.35~Mcps\\
$[$ @ 25$^\circ$C $]$  & ($>$1.2~pe) & $<$0.135~Mcps \\
\hline \hline
\end{tabular}
\label{tab:spec}
\end{center}
\end{table}%

The production of MPPCs was started in February 2008, and finished in February 2009.
In total about 64,500 MPPCs including spares were produced and delivered in about a year.
Among them, 17,686 MPPCs were tested at Kyoto University,while remainder was tested at other collaborating institutes.
One of such measurements is described in~\cite{LLR-MPPC}.
In the following sections, we describe the test system and measurement results at Kyoto university.

\section{Measurement setup}

\subsection{Overview of measurement}
In order to characterize a large number of MPPCs, we have developed a test system that simultaneously measures 64 MPPCs.
Figure~\ref{fig:mppc_test_bench_kyoto} shows the schematics of the test system.
We recorded the charge from the MPPCs with and without external light, for a range of the operation voltage with 0.1~V step, and with three temperature settings at 15, 20, and 25\degC.
From those data, we are able to extract the gain, dark noise rate, PDE, and cross-talk and after-pulsing probability of each MPPC as functions of the operation voltage and temperature.
Based on the measurement result, we have checked if the MPPC satisfies our requirements.
In addition, those data will be used as the reference after MPPCs are installed into the detector, although the final calibration will be done \textit{in-situ} using data taken during the experiment.

\subsection{MPPC test system}

\begin{figure}[!t]
\begin{center}
\centering
\includegraphics[width=0.45\textwidth]{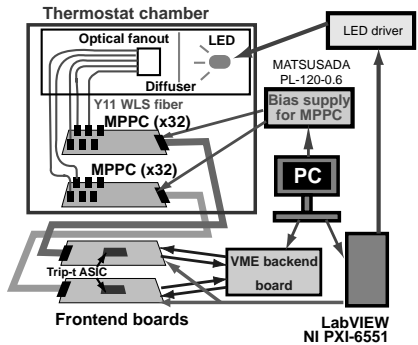}
\caption{Schematic diagram of the MPPC test system.
Sixty four MPPCs are simultaneously tested.}
\label{fig:mppc_test_bench_kyoto}
\end{center}
\end{figure}

Figure~\ref{fig:mppc_test_bench_kyoto} shows a schematic diagram of the MPPC test system.
It consists of a light source (LED), readout electronics, a thermostatic chamber and a control system.

The light from a blue LED (Nichia NSPB500S) is diffused with a set of plastic plates and distributed to WLS fibers via an optical fanout.
The other end of the WLS fiber is connected to an MPPC using the optical connector developed for T2K detectors~\cite{pd07_kawamuko}.
Kuraray Y11(200)MS WLS fiber with 1 mm diameter, the same type as we use in the real T2K detectors, is used in this test system.
The light intensity at the end of each fiber is measured with a PMT before the mass measurement.

The signal from the MPPC is read out by the TriP-t ASIC developed at Fermilab~\cite{Trip-t}, which is also used for the T2K near detectors.
We use the TriP-t chip to integrate the charge within the gate timing and to serialize output from MPPCs.
The TriP-t has 32 input channels and we can measure 64 MPPCs at one time using two TriP-t chips.
The charge of each MPPC is recorded with VME-based electronics developed by our group \cite{pd07_murakami}.
The digital signal to control TriP-t and LED pulsing is generated using National Instruments PXI-6551.
The bias voltage for MPPCs and VME readout are controlled by a Linux PC.

\begin{figure}[!t]
\begin{center}
\centering
\includegraphics[width=0.45\textwidth]{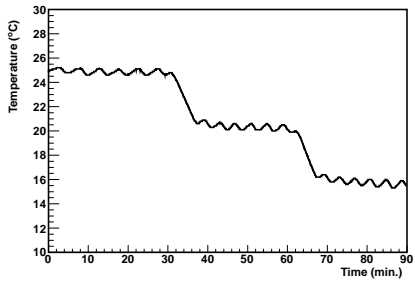}
\caption{The temperature variation in the thermostat chamber during the measurement.
}
\label{fig:temp_thermo_chamber}
\end{center}
\end{figure}

The light source and MPPCs are kept inside a lightproof thermostatic chamber.
The temperature inside the box is automatically controlled and changed to 25, 20 and 15\degC, in this order, every 30 minutes after starting the test sequence.
Thus, data at three temperatures are automatically taken in 1.5 hours.
A thermometer (T\&D TR-71U) is installed inside the chamber to monitor the temperature during the measurement.
Figure~\ref{fig:temp_thermo_chamber} shows the temperature variation in the chamber during the measurement.
The temperature becomes stable at each measurement point after about 10 minutes.
The measurement is performed in the following 20 minutes.
There is a short-period temperature oscillation with $\simeq 0.4$~\degC\ amplitude and 4 minutes cycle caused as the result of the temperature control by the chamber.
It corresponds to a variation of $\pm$20~mV in \Vbd\ and consequently in \delV.
Which gives 2\% uncertainty when $\delV=1$~V.

During the measurement, the overvoltage $\delV$ was scanned from 0.7 V to 1.8 V in steps of 0.1 V.
At each voltage, 8,000 events were taken with the light injection and with a 200 ns ADC gate,
and other 8,000 events were taken without external light and with an 800 ns ADC gate (for the noise rate measurement).
This procedure was repeated at 15, 20 and 25\degC.

\section{Data analysis and performance of T2K-MPPC}
The methods to derive the MPPC parameters from the ADC distribution are given in the following sections, together with measurement results.
The measured parameters for 17686 MPPCs  at 15, 20, 25 \degC \ and \delV = 1.0 %and 1.4~$V$ 
are summarized in Table~\ref{tab:summary_measurement}.

\subsection{Failure rate}
We found that nine MPPCs, out of 17,695, did not return signal with the bias voltage in the scanned range.
A few MPPCs were further tested and found to show no current flow even with a forward bias, indicating some part of circuit inside MPPC is broken.
They were working without problems before shipment according to the test sheet from Hamamatsu Photonics.
The reason of failure is under investigation but not yet known.
All the remaining MPPCs are confirmed to satisfy our requirements.
Thus, only 0.05\% of delivered device was rejected.

\subsection{Gain and Breakdown voltage}
\begin{figure}[!t]
\begin{center}
\centering
\includegraphics[scale=0.35, angle=0]{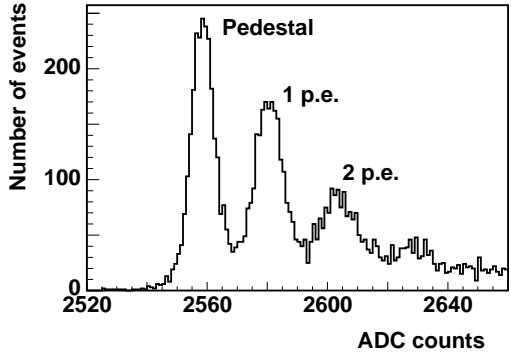}
\caption{ADC distribution of MPPC with LED on.}
\label{fig:adc_distribution}
\end{center}
\end{figure}

\begin{figure}[!t]
\begin{center}
\centering
\includegraphics[scale=0.35, angle=0]{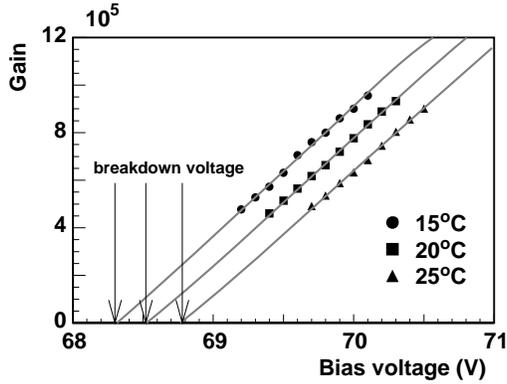}
\caption{Determination of \Vbd\ from gain-voltage extrapolation.
Circles, squares, and triangles are for data at 15, 20, 25\degC,
respectively.}
\label{fig:breakdown_voltage}
\end{center}
\end{figure}

The gain of MPPC can be easily measured because the pedestal and one photoelectron (p.e.) peak are well separated (Fig.~\ref{fig:adc_distribution}).
Measuring the charge corresponding to one p.e.\ and dividing it by the electron charge, we can obtain the gain.
The gain changes linearly on \delV, as shown in Eq.~\ref{eq:gain-Vdep}.
We can derive $\Vbd$ by linearly extrapolating the gain-voltage relation to the point where gain becomes zero (Fig.~\ref{fig:breakdown_voltage}).

\begin{figure}[!t]
\begin{center}
\centering
\includegraphics[scale=0.35, angle=0]{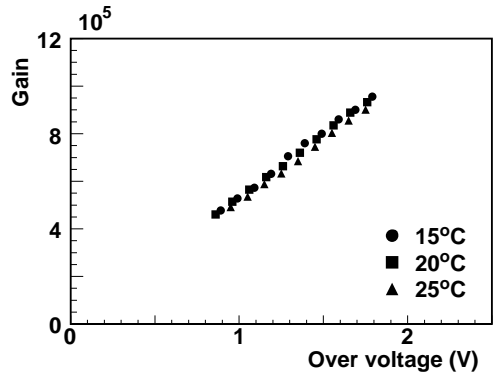}
\caption{Measured gain as a function of \delV\ for one MPPC.
Circles, squares and triangles are for data at 15, 20, 25\degC, respectively.}
\label{fig:overvoltage_gain}
\end{center}
\end{figure}

Figure~\ref{fig:overvoltage_gain} shows the measured gain as a function of $\delV$ at 15, 20, 25\degC\ for one MPPC.
The gain is independent of the temperature if \delV\ is the same, although \Vbd\ is dependent on the temperature.

\begin{figure}[!t]
\begin{center}
\centering
\includegraphics[scale=0.35, angle=0]{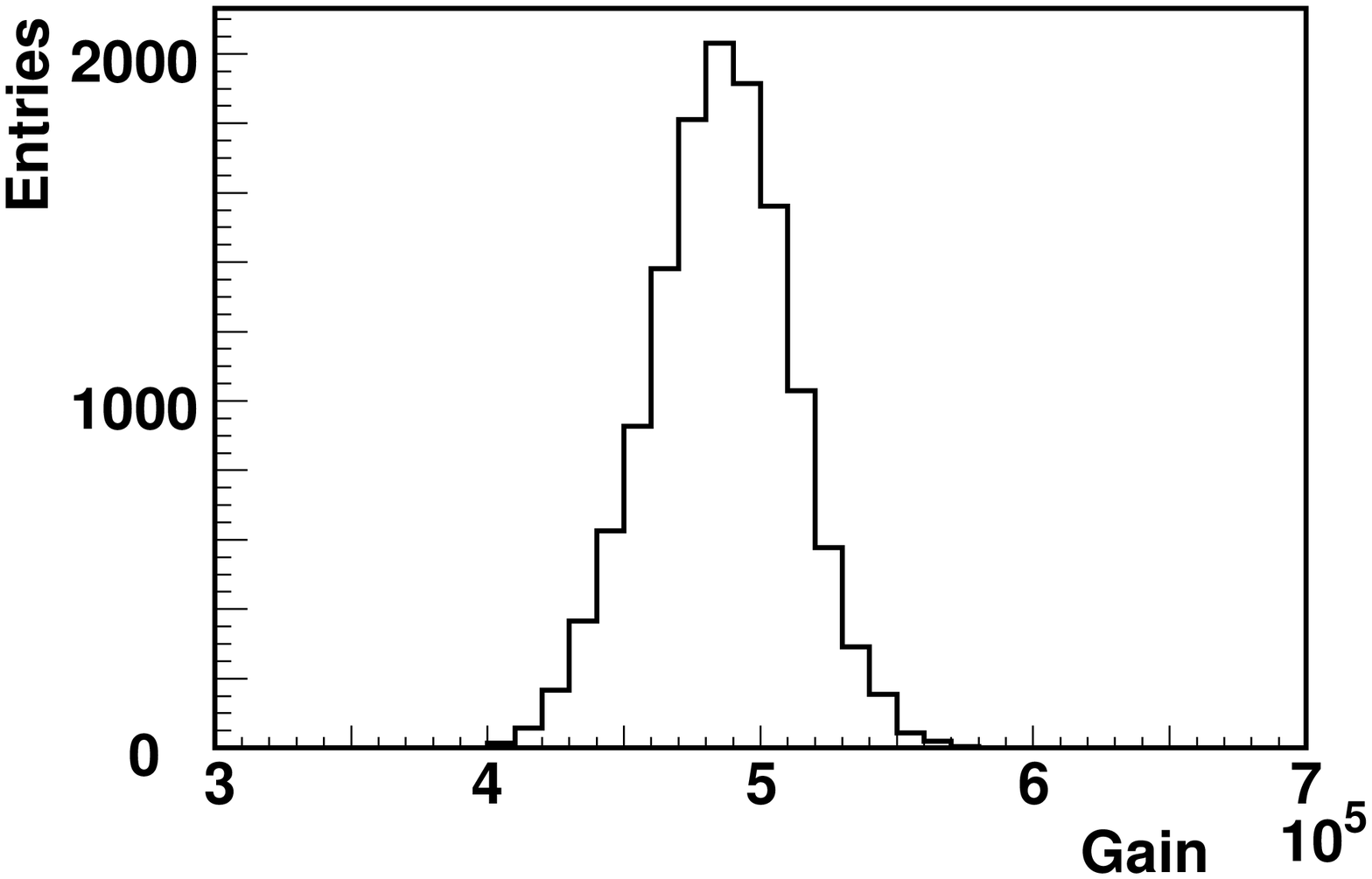}
\caption{Gain at $\delV=1.0$~$V$ and 20\degC \ for 17686 MPPCs.}
\label{fig:gain_dv10}
\end{center}
\end{figure}

Figure~\ref{fig:gain_dv10} shows the distribution of measured gain at $\delV =$ 1.0~$V$ and 20\degC\ 
for 17686 MPPCs.
The average gain is measured to be about $4.85 \times 10^{5}$ at  $\delV =$ 1.0~$V$ and 20\degC.
The root mean square (RMS) of the distribution is 5.4~\%,
while the measurement systematics, dominated by the uncertainty in the inter-channel gain calibration of the readout electronics, is estimated to be around 4\%.

\begin{figure}[!t]
\begin{center}
\centering
\includegraphics[scale=0.35, angle=0]{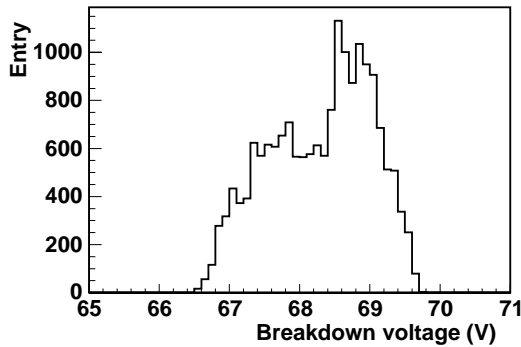}
\caption{Measured breakdown voltage at 20\degC\ for 17686 MPPCs.}
\label{fig:vbd_serial}
\end{center}
\end{figure}

Figure~\ref{fig:vbd_serial} shows the measured breakdown voltage \Vbd\ at 20\degC. 
The breakdown voltage is in the range of 66.5--69.7~$V$.
Before shipment, Hamamatsu has measured the voltage which gives a gain of $7.5\times10^5$ at 25\degC\ for each MPPC.
It is confirmed that our measurement is consistent with Hamamatsu test sheet, except for a systematic shift of the voltage presumably due to the difference of the measurement system.

As already mentioned earlier, \Vbd\ depends on the temperature.
The temperature coefficient of \Vbd\ is measured to be 48~mV/\degC\ in our system.

\subsection{Dark noise rate}
From the data without light source, dark noise rate is measured by counting the number of dark events and dividing it by the measurement time.
We assume that the number of true dark events, in absence of the effect from cross-talk and after pulsing, follows Poisson statistics.
In that case, the average number of true dark events $n_\mathrm{dark}$ can be estimated from the fraction of pedestal events ($n^\mathrm{obs}_\mathrm{0pe}$) among total events ($N$), $P(0) \equiv n^\mathrm{obs}_\mathrm{0pe} / N$, as

\begin{equation}
n_\mathrm{dark} = -\ln\left( P(0) \right).
\end{equation}
Dividing $n_\mathrm{dark}$ by the gate width, the dark noise rate is calculated.

\begin{figure}[!t]
\begin{center}
\centering
\includegraphics[scale=0.35, angle=0]{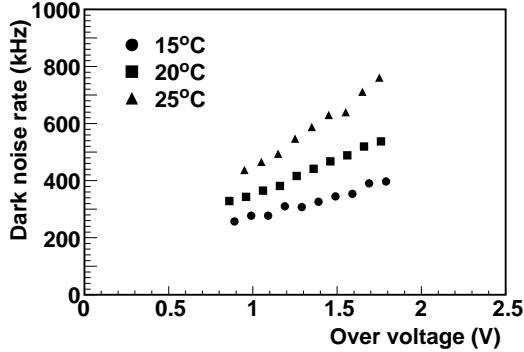}
\caption{Measured dark rate as a function of \delV\ for one MPPC.
Circles, squares and triangles are for data at 15, 20, 25\degC, respectively.}
\label{fig:overvoltage_dark}
\end{center}
\end{figure}

Figure~\ref{fig:overvoltage_dark} shows the dark noise rate as a function of \delV\ at 15, 20, 25\degC\ for one MPPC.
Unlike other parameters, even at the same \delV, the dark rate decreases as the temperature becomes lower 
due to the strong temperature dependence of the number of thermally generated carriers that are the origin of dark noise.
%%%

\begin{figure}[!t]
\begin{center}
\centering
\includegraphics[scale=0.35, angle=0]{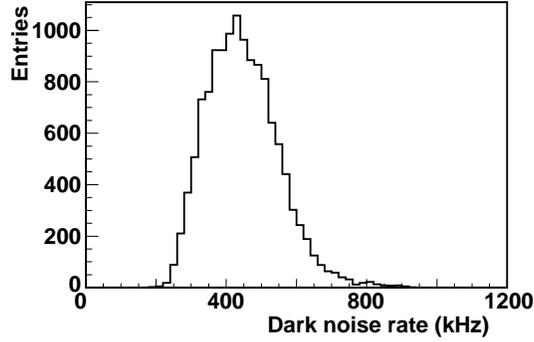}
\caption{Dark rate at $\delV =$ 1.0~$V$ and 20\degC\  for 17686 MPPCs.}
\label{fig:dark_dv10}
\end{center}
\end{figure}

Figure~\ref{fig:dark_dv10} shows the distribution of measured dark noise rate at $\delV =$ 1.0~$V$ and 20\degC\  for 17686 MPPCs.
The average dark rate is measured to be about 0.45 MHz at $\Delta {\rm V} = 1.0\ {\rm V}$ and 20\degC.
The RMS of the distribution is around 25~\%,
while the measurement systematics, coming from the temperature variation during the measurement and uncertainty of estimating $n^\mathrm{obs}_\mathrm{0pe}$, is estimated to be around 5\%.
Thus, the dark noise rate has large device by device variation.

\subsection{After pulsing and cross-talk probability}
MPPCs are known to exhibit correlated noise due to cross-talk and after-pulsing as described earlier.
A dark noise, or photo-electron triggered avalanche may indeed yield additional avalanches due to both phenomena hence increasing the total detected charge.
Although the after pulsing has a different timing structure from cross-talk, only the sum of two effects is estimated with measuring the charge integrated in a gate time window, which is sufficient for the purpose of quality assurance for the application to T2K ND280.

We can estimate the true number of one p.e.\ events $n^\mathrm{true}_\mathrm{1pe}$, without the effect from after pulsing and cross-talk, from the fraction of pedestal events and Poisson statistics as is done in the dark noise rate measurement.
In reality, the number of one p.e.\ events is less than  $n^\mathrm{true}_\mathrm{1pe}$ due to the effect of after pulsing and cross-talk.
Comparing  $n^\mathrm{true}_\mathrm{1pe}$ with the observed number of events at one p.e.\ peak $n^\mathrm{obs}_\mathrm{1pe}$, we estimate the probability of after pulsing and cross-talk $p_\mathrm{apct}$ as
\begin{equation}
p_\mathrm{apct} = 1-\frac{n^\mathrm{obs}_\mathrm{1pe}}{n^\mathrm{true}_\mathrm{1pe}}.
\end{equation}

\begin{figure}[!t]
\begin{center}
\centering
\includegraphics[scale=0.35, angle=0]{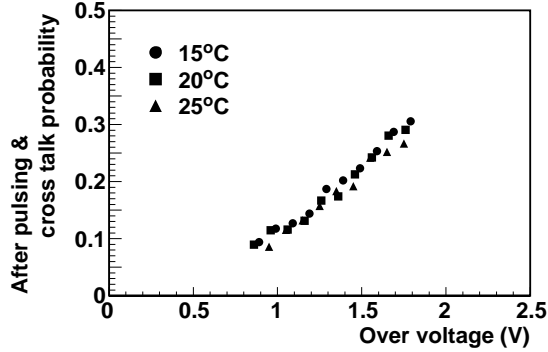}
\caption{Measured after pulsing and cross-talk probability as a function of \delV\ for one MPPC.
Circles, squares and triangles are for data at 15, 20, 25\degC, respectively.}
\label{fig:overvoltage_apct}
\end{center}
\end{figure}

Figure~\ref{fig:overvoltage_apct} shows the after pulsing and cross-talk probability
as a function of the over voltage \delV\ at 15, 20, 25\degC \ for one MPPC.
The after pulsing and optical cross-talk probability is independent of the temperature if \delV\ is kept the same.

\begin{figure}[!t]
\begin{center}
\centering
\includegraphics[scale=0.35, angle=0]{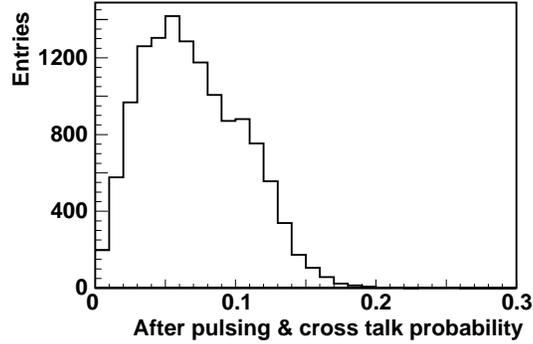}
\caption{After pulsing and cross talk probability at \delV = 1.0~$V$ and 20\degC \  for 17686 MPPCs.}
\label{fig:apct_dv10}
\end{center}
\end{figure}

Figure~\ref{fig:apct_dv10} shows the distribution of measured after pulsing and cross-talk probability
at  $\delV =$ 1.0~$V$ and 20\degC\ for 17686 MPPCs.
The after pulsing and cross-talk probability of MPPC is measured to be about 0.07 on the average at $\delV =$ 1.0~$V$ and 20\degC.
The systematic variation in the measurement, mainly coming from the uncertainty in estimating $n^\mathrm{obs}_\mathrm{1pe}$, is estimated to be about 0.04, while the RMS of the distribution in Fig.~\ref{fig:apct_dv10} is 0.036.
Thus, the variation of after-pulse and cross-talk probability seen in Fig.~\ref{fig:apct_dv10} is dominated by the measurement systematics and the device uniformity itself is considered to be much better.

\subsection{Photon detection efficiency (PDE)}
\begin{figure}[!t]
\begin{center}
\centering
\includegraphics[scale=0.35, angle=0]{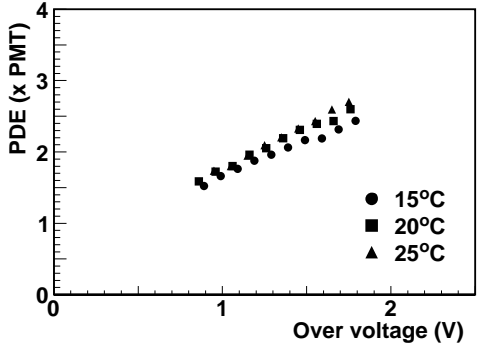}
\caption{Measured PDE as a function of \delV\ for one MPPC.
Circles, squares and triangles are for data at 15, 20, 25\degC, respectively.}
\label{fig:overvoltage_pde}
\end{center}
\end{figure}

Because measuring the absolute value of the photon detection efficiency (PDE) is difficult in the mass measurement, it is relatively measured using a PMT (Hamamatsu R6427) as a reference.
Referring to the Hamamatsu datasheet, the quantum efficiency of the reference PMT is 10--15 \% for the wavelength of 470-530~nm, which is the range of the emission light from the Y11 WLS fiber.
In order to avoid the effect of after pulsing and cross-talk, the number of photoelectrons detected with each MPPC is derived from the fraction of pedestal events in the same way as explained above, in the presence of weak light source.
The effect of non-uniform light distribution among 64 WLS fibers are corrected by the measurement of light intensity with the PMT.
The stability of the LED light intensity is monitored with a reference MPPC.
The PDE is defined as the number of photoelectrons detected by MPPC divided by that of the PMT using the same 1~mm diameter WLS fiber.
Figure~\ref{fig:overvoltage_pde} shows the measured PDE as a function of \delV\ at 15, 20, 25\degC \ for one MPPC.
The PDE is independent of temperature if \delV\ is kept the same.

\begin{figure}[!t]
\begin{center}
\centering
\includegraphics[scale=0.35, angle=0]{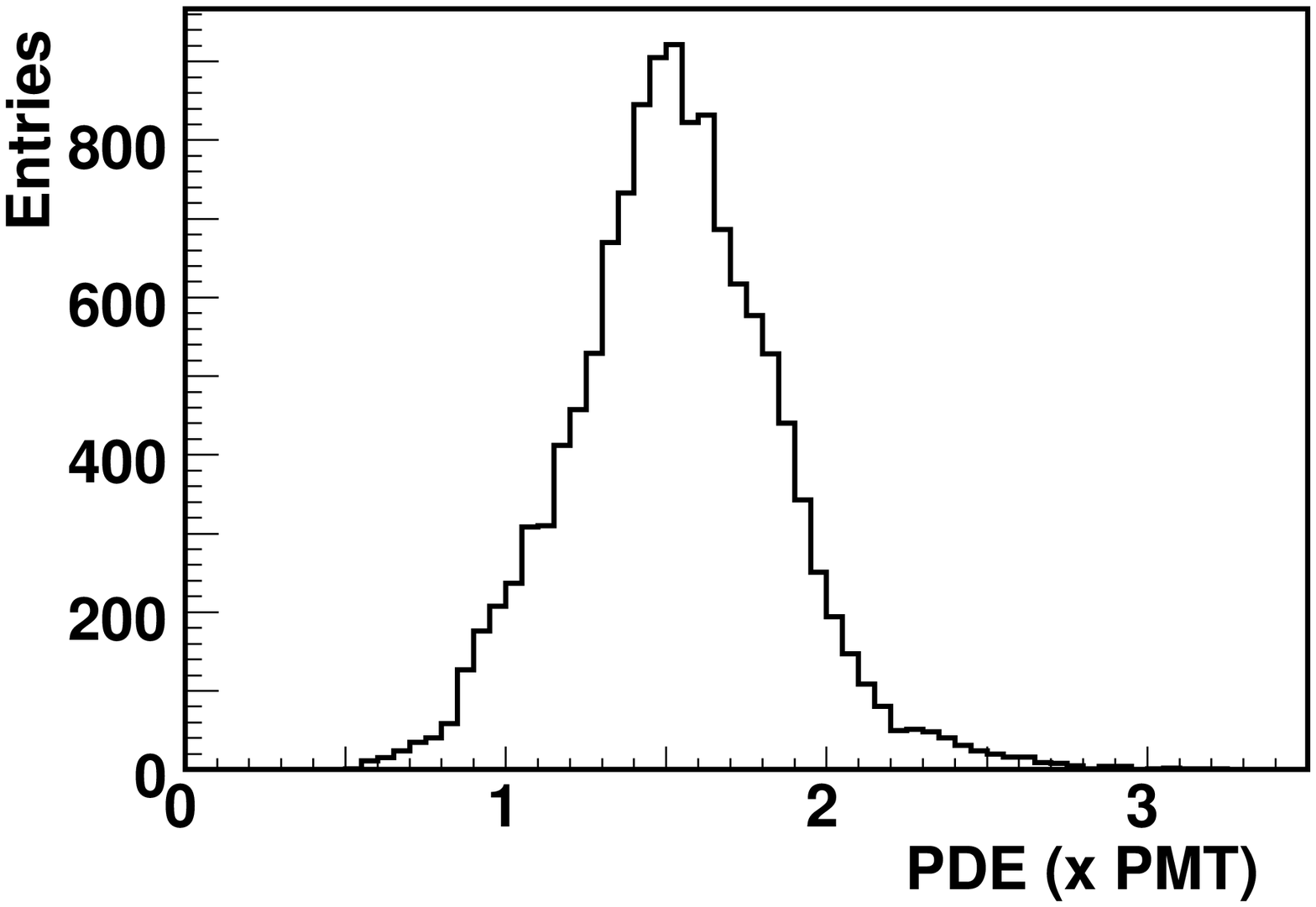}
\caption{PDE at \delV = 1.0~$V$ and 20\degC \ for 17686 MPPCs.}
\label{fig:pde_dv10}
\end{center}
\end{figure}

Figure~\ref{fig:pde_dv10} shows the distribution of PDE at \delV= 1.0~$V$ and 20\degC\ for 17686 MPPCs.
The PDE of MPPC is measured to be about 1.5 times PMT for the green light from WLF fiber Y11(200) at \delV= 1.0~$V$ and 20\degC.
The systematic uncertainty in the measurement is estimated to be about 20\%, dominated by the non-uniformity and reproducibility of the light distribution to each MPPC.
The RMS of the distribution is 22\%.
Thus, the variation is dominated by the systematic uncertainty of the measurement.
The tail seen in the higher part of the distribution is also considered to be due to the systematics of the measurement.

\subsection{Performance summary}
Table~\ref{tab:summary_measurement} summarizes  the mean value and RMS (in parentheses) of measured MPPC performance for 17,686 MPPCs  at 15, 20, 25 \degC \ and \delV = 1.0. %and 1.4~$V$.
All the MPPCs satisfy the requirement to be used in the T2K near neutrino detectors.
At a fixed \delV, slight dependence on temperature is seen in the gain, PDE and afterpulsing/cross-talk probability.
However, more accurate, independent measurement of several MPPCs reveals no such dependence.
Therefore, although we have not fully understood the reason, it is presumably due to the feature of our measurement system and not the performance of the MPPC.

\begin{table}[h]
 \begin{center}
 \begin{tabular}{cccc}
 \hline \hline
Parameter  & Temperature & Measured values & RMS \\
 \hline

		& 15\degC & $4.91 \times 10^{5}$ & $0.26\times 10^{5}$ \\
Gain 	& 20\degC & $4.85 \times 10^{5}$ & $0.26\times 10^{5}$  \\
		& 25\degC & $4.75 \times 10^{5}$ & $0.24\times 10^{5}$ \\
\hline
Breakdown voltage 	& 15\degC & 68.05 &  0.73 \\
(V)         			& 20\degC & 68.29 &  0.73 \\
                  		& 25\degC & 68.53 &  0.73 \\
\hline
Dark noise rate 	& 15\degC & $3.37 \times 10^{5}$ &  $0.85\times 10^{5}$ \\
(Hz)      			& 20\degC & $4.47 \times 10^{5}$ &  $1.02\times 10^{5}$ \\
               			& 25\degC & $6.03 \times 10^{5}$ &  $1.21\times 10^{5}$  \\
\hline
After pulsing and 	& 15\degC & 0.073 &  0.039  \\
cross-talk       		& 20\degC & 0.070 &  0.036  \\
probability      		& 25\degC & 0.066 &  0.031  \\
\hline
Relative PDE 		& 15\degC & 1.45 &  0.32 \\
($\times$PMT)     	& 20\degC & 1.53 &  0.33 \\
                            	& 25\degC & 1.62 &  0.34 \\
 \hline
 \end{tabular}
 \end{center}
 \caption{Mean value and RMS of gain, dark noise rate,
after pulsing and cross-talk probability and photo detection efficiency
for 17686 MPPCs 
at 15, 20, 25 \degC \ and \delV= 1.0~$V$.}
 \label{tab:summary_measurement}
\end{table}

\section{Summary}
For the T2K near neutrino detectors, we have developed the Multi-Pixel Photon Counter (MPPC) with 1.3$\times$1.3~mm$^2$ active area and with 667 50~$\mu$m pitch pixels (S10362-13-050C).
In total, about 64,500 MPPCs were produced for T2K from February 2008 to February 2009. 

At Kyoto University, we have successfully developed the system and technique to characterize a large number of MPPCs, and tested 17,686 MPPCs.
The performance of MPPCs has been confirmed to satisfy our requirements.
The failure rate is found to be about 0.5\%.

\section*{Acknowledgments}
The authors are grateful to the solid state division of Hamamatsu Photonics for providing us test samples during the development.
We appreciate useful suggestions from the members of T2K ND280 photosensor group.
We thank F.~Retiere for careful reading and helpful comments on this manuscript. 
This work was supported by MEXT and JSPS with the Grant-in-Aid
for Scientific Research A 19204026, Young Scientists S 20674004,
Scientific Research on Priority Areas ``New Developments of Flavor Physics'', 
and the global COE program ``The Next Generation of Physics, Spun from Universality and Emergence''.
The development of MPPC and the test system is also supported by the KEK detector technology project.

%% The Appendices part is started with the command \appendix;
%% appendix sections are then done as normal sections
%% \appendix

%% \section{}
%% \label{}


\begin{thebibliography}{00}

%% \bibitem{label}
%% Text of bibliographic item

\bibitem{T2K} Y.~Itow et al.,  %[The T2K Collaboration],
  ``The JHF-Kamioka neutrino project,''
 hep-ex/0106019.

\bibitem{T2K-ND280}
  Y.~Kudenko  [T2K Collaboration],
  %``The near neutrino detector for the T2K experiment,''
  Nucl.\ Instr.\ and Meth.\  A {\bf 598}, 289 (2009)
  [arXiv:0805.0411 [physics.ins-det]].
  %%CITATION = NUIMA,A598,289;%%

\bibitem{sci-wls} 
A.~Pla-Dalmau, %[MINOS Scintillator Group],
Frascati Phys.\ Ser.\  {\bf 21}, 513 (2001);  % MINOS
K.~Nitta et al., Nucl.\ Instr.\ and Meth.\  A {\bf 535}, 147 (2004); % K2K
D.~Drakoulakos et al., FERMILAB-PROPOSAL-0938, arXiv:hep-ex/0405002; % MINERvA
A.~A.~Aguilar-Arevalo et al., FERMILAB-PROPOSAL-0954, arXiv:hep-ex/0601022; % SciBooNE 
T.~Adam et al., Nucl.\ Instr.\ and Meth.\  A {\bf 577}, 523 (2007). % OPERA

\bibitem{MPPC}
 M.~Yokoyama et al., arXiv:physics/0605241;
  S.~Gomi et al., PoS {\bf PD07}, 015 (2007).

\bibitem{INGRID}
M.~Otani, PoS {\bf PD09}, 020 (2009).

\bibitem{FGD}
K.~Ieki, PoS {\bf PD09}, 023 (2009).

\bibitem{HPK} %MPPC references.
Hamamatsu Photonics K.\ K.,
 [Online]: http://www.hamamatsu.com .

\bibitem{Otono:2006zz}
  H.~Otono, H.~Oide, T.~Suehiro, H.~Hano, S.~Yamashita and T.~Yoshioka,
  %``Study of MPPC at liquid nitrogen temperature,''
  PoS {\bf PD07}, 007 (2006).


\bibitem{taguchi-mthesis}
M.~Taguchi, ``Development of Multi-Pixel Photon Counters and readout electronics'', Master's thesis, Kyoto University (2006).

\bibitem{crosstalk-ref}
N.~Akil {\it et al.},
``A multimechanism model for photon generation by silicon junctions in avalanche breakdown,''
IEEE Trans. Electron Devices, \textbf{46} 1022 (1999).

\bibitem{Oide}
H.~Oide, T.~Murase, H.~Otono and S.~Yamashita,
%``Studies on multiplication effect of noises of PPDs, and a proposal of a new
%structure to improve the performance,''
Nucl.\ Instrum.\ Meth.\  A {\bf 613}, 23 (2010).
% arXiv:0811.1402 [physics.ins-det].
%%CITATION = ARXIV:0811.1402;%%

\bibitem{LLR-MPPC}
F.~Moreau, J.-C.~Vanel, O.~Drapier, M.~Gonin, A.~Bonnemaison, A.~Cauchois, Y.~Geerebaert, S.~Couturier-Le Quellec,
Nucl.\ Instrum.\ Meth.\  A {\bf 613}, 46 (2010).


\bibitem{pd07_kawamuko}
H. Kawamuko, T. Nakaya, K. Nitta and M. Yokoyama, PoS {\bf PD07}, 043 (2007).

\bibitem{Trip-t}
J.~Estrada, C.~Garcia, B.~Hoeneisen and P.~Rubinov,
FERMILAB-TM-2226 (2003).

\bibitem{pd07_murakami}
T. Murakami and S. Gomi, PoS {\bf PD07}, 046 (2007).

\end{thebibliography}
\end{document}